\documentclass[preprint,showpacs,preprintnumbers,eqsecnum]{revtex4}
\usepackage{graphicx}
\usepackage{dcolumn}
\usepackage{bm}

\def\lesssim{\ \raise.3ex\hbox{$<$}\kern-0.8em\lower.7ex\hbox{$\sim$}\ }
\def\gesim{\ \raise.3ex\hbox{$>$}\kern-0.8em\lower.7ex\hbox{$\sim$}\ }
\font\scripti=cmmi7
\font\scriptscripti=cmmi5
\def\sib#1{\setbox0 = \hbox{\scripti #1}
  \kern-.02em\copy0\kern-\wd0
  \kern.04em\box0} 
\def\ssib#1{\setbox0 = \hbox{\scriptscripti #1}
  \kern-.02em\copy0\kern-\wd0
  \kern.04em\box0} 
\font\tenib=cmmib10 
\skewchar\tenib='177 \skewchar\tenib='177 \skewchar\tenib='177
\def\pbold#1{\setbox0 = \hbox{$ #1 $}
  \kern-.022em\copy0\kern-\wd0
  \kern.011em\copy0\kern-\wd0
  \kern.011em\copy0\kern-\wd0
  \kern.011em\copy0\kern-\wd0
  \kern.011em\box0} 
\begin{document}
\title{Collective modes and the effect of single-particle excitations in 
the BCS-BEC crossover region of a trapped Fermi superfluid}
\author{Y. Ohashi$^{1,2}$, and A. Griffin$^2$}
\affiliation{$^1$Institute of Physics, University of Tsukuba, Tsukuba,
  Ibaraki 305, Japan, \\
$^2$Department of Physics, University of Toronto, Toronto,
  Ontario, Canada M5S 1A7}
\date{\today}
\begin{abstract}
We investigate the collective oscillations in the BCS-BEC crossover region 
of a trapped Fermi superfluid at $T=0$. We explicitly include 
the effect of a Feshbach resonance, which leads to a tunable pairing interaction
adjusted by the threshold energy of the resonance, as well as the associated molecules.
In a previous paper,we obtained solutions of the 
Bogoliubov-de Gennes coupled equations, describing a trapped Fermi superfluid gas.
Using these Bogoliubov quasi-particle excitations, we 
calculate the density correlation function in 
the Hartree-Fock-Bogoliubov generalized random phase approximation
(HFB-GRPA).
Previous discussions of collective modes 
were based on solving the $T=0$ equation of motion for the 
superfluid in a trap, ignoring the effect of single-particle excitations.
We determine the frequencies of the quadrupole and the monopole modes.
In the crossover region, the frequencies of these two modes are shown to be strongly 
depressed by the threshold energy of the two-particle continuum, 
originating from Andreev single-particle 
bound states (or in-gap states) localized
at the edge of the trap. 
This effect becomes less important as one enters the BEC regime, due to 
the absence of these low energy fermion bound states.
This suppression of collective mode frequencies by the two-particle continuum
 may be a useful experimental signature of the single-particle energy
gap in the crossover region of trapped Fermi superfluids.
\end{abstract}
\pacs{03.75.Ss, 03.75.Kk, 03.70.+k}
\maketitle
%
\section{Introduction}
The recently discovered fermion superfluidity in $^{40}$K\cite{Jin} and $^6$Li\cite{Bartenstein,Zwierlein,Kinast,Bourdel} has attracted much attention. This is the first system where we can experimentally examine the BCS-BEC crossover phenomenon\cite{Ohashi1,Ohashi2,Ohashi3,Milstein,Perali}, originally discussed in the superconductivity literature\cite{Eagles,Leggett,Nozieres,Tokumitsu,Randeria,Melo,Engelbrecht,Haussmann}. The novel pairing mechanism of this Fermi superfluid is associated with a Feshbach resonance\cite{Ohashi1,Timmermans,Holland}, in which two Fermi atoms can form a quasi-molecular (boson) state. The resulting attractive interaction can be tuned by varying the threshold energy (denoted in this paper by $2\nu$) of the resonance, using an external magnetic field\cite{Jin2,Ketterle}. 
\par
In Fermi superfluids, there are two distinct kinds of excitations, i.e., collective modes and single-particle excitations. Two single-particle excitations are created with the breakup of a Cooper-pair. Even in the BEC regime, where the character of superfluidity can be best described as the Bose condensate of molecules, the single-particle Fermi excitations still exist. This situation quite contrasts to an ordinary BEC of Bose atoms, where low-energy excitations are all bosonic in nature\cite{Griffin}. Very recently, both the single-particle excitations\cite{Chin} and collective modes\cite{Kinast,Bar2} have been studied experimentally in the crossover region of superfluid $^6$Li.
\par
The frequency of a collective mode is strongly affected when it is near the threshold energy of the two-particle continuum at $2E_g$, where $E_g$ is the single-particle excitation gap. In the BCS region of a uniform, neutral Fermi superfluid, this can be seen by the suppression of the Goldstone mode ($\omega={v_{\rm F} \over \sqrt{3}}q$ at low $q$), when its energy approaches the two-particle continuum value of $2|\Delta|$\cite{OhashiT}. 
Here, $v_F$ is the Fermi velocity and $\Delta$ is the BCS Cooper-pair order parameter. 
A similar suppression of a collective mode frequency 
by the single-particle excitations is expected in a trapped Fermi superfluid. In this case, we have shown recently\cite{Ohashi5} that the single-particle energy gap $E_g$ is {\it not} simply related to the position-dependent order parameter $\Delta({\bf r})$, but is dominated by the lowest energy Andreev bound state appearing at the edge of the trapped gas\cite{Baranov,Saint}. Even in the crossover region, where the magnitude of the superfluid order parameter in the center of the trap is comparable to the Fermi energy $\varepsilon_{\rm F}$, $E_g$ is still the order of the trap frequency $\omega_0$ (which is usually much smaller than $\varepsilon_{\rm F}$). Since the frequencies of the quadrupole and the monopole modes are the order of $\omega_0$\cite{Pethick}, they are expected to be strongly affected by the two-particle continuum at $2E_g$. 
\par
In this paper, we investigate the collective modes at $T=0$ in the BCS-BEC crossover region of a Fermi superfluid in a trap. 
The theory is precisely the same as in our recent discussion of collective modes in a uniform Fermi superfluid\cite{Ohashi3}, except that we limit ourselves to $T=0$.
The present paper makes crucial use of our results\cite{Ohashi5} for the single-particle excitations in a trapped Fermi gas superfluid.
To properly include the effect of the energy gap $E_g$ associated with the Andreev bound states, the self-consistent Bogoliubov de Gennes (BdG) coupled equations\cite{BdG} were solved numerically. The strong-coupling effect is taken into account by considering the deviation of the chemical potential $\mu$ from the Fermi energy $\varepsilon_{\rm F}$, which is associated with the appearance of the bosonic pair states. As a result, the BdG equations have to be solved together with the equation of state involving the number of atoms\cite{Ohashi5} to take into account the molecule condensation.
\par
In this paper, the single-particle solutions of the BdG equations are used in a detailed calculation of the density correlation function, using the time-dependent Hartree-Fock-Bogoliubov (random phase) approximation (HFB-GRPA). This includes the coupling between ordinary density fluctuations and fluctuations in the Cooper pair-channel. 
As proven recently\cite{Ohashi4}, this linear response formalism guarantees that Goldstone's theorem and Kohn's theorem\cite{Kohn} are satisfied in the entire BCS-BEC crossover regime, required for any consistent approximation for collective modes in a Fermi superfluid.
\par
Our approach is an extension of the weak-coupling BCS theory by Bruun\cite{Bruun1,Bruun2} to the strong-coupling regime including a Feshbach resonance and the associated molecules. In the crossover region, we show that the frequencies of the low-energy collective excitations, such as the quadrupole and monopole modes, are strongly suppressed so that they stay below the threshold energy of the two-particle continuum at $2E_g$. We note that the $T=0$ hydrodynamic superfluid theory\cite{Pethick,Stringari,Pita,Pita2} describing pure superfluid dynamics cannot, in principle, describe this suppression phenomenon. This is because such a theory ignores any explicit consideration of single-particle excitations, and in particular the low energy Andreev excitations in a trap. As expected, this suppression effect disappears as one enters the BEC region, where the single-particle excitation energy gap becomes very large\cite{Ohashi5}. 
\par
It is useful to be clear about the relation of our kind of microscopic theory of collective modes and the predictions based on solving the equation of motion of a pure superfluid\cite{Pethick,Pita,Pita2}. The latter so-called ``hydrodynamic'' theory is limited to $T=0$, since it assumes that all the atoms are described by the superfluid component (i.e., there is no normal fluid). In this theory, the only unknown microscopic input is the equilibrium equation of state, giving the chemical potential as a function of the density. There has been considerable recent progress in determining this equation of state in the BCS-BEC crossover regime, including the strong interaction unitarity limit ($(k_{\rm F}a_{\rm F})^{-1}=0$)\cite{Gio7,Hei7}. The advantage of starting from the equation of motion for a pure superfluid is that it is expected to be valid even in the strong interaction limit, and makes no assumption about the amount of depletion of the condensate (which could be very large in the unitarity limit). The weakness of the $T=0$ superfluid hydrodynamic theory is that, by definition, it cannot deal with any questions related to the single-particle excitations, such as the effect of the two-particle continuum or the effect of a normal fluid component (which would arise at finite temperatures). In contrast, our present microscopic calculations can deal with such effects. However, they do not include any large depletion of the condensate due to interaction, although the effect of thermal excitations can be treated.
\par
One deficiency of our present calculations might be mentioned at the outset, namely that in the BEC limit, our collective modes are predicted to be those of a non-interacting molecular Bose condensate. The explanation of this is simply that in mean-field BCS theory which lies at the core of our present discussion, we do not include the higher order fluctuations which ultimately give rise to the correct $s$-wave interaction between molecules $a_{\rm M}=0.6a_{\rm F}$\cite{Petrov7}, in the BEC limit of the crossover. In a separate paper\cite{Ohashi7}, we give the appropriate extension which does reproduce the correct value of $a_{\rm M}$, which in turn would lead to correct collective mode frequencies in the BEC limit\cite{Stringari2,note7}.
However, this deficiency in reproducing the BEC collective modes correctly is of minor concern in connection with the major point we emphasize in this paper, namely the suppression of the collective mode frequencies in the BCS-BEC crossover region due to coupling to the two-particle continuum.
\par
It should also be emphasized that our microscopic HFB-GRPA
theory for the collective modes could be used as the starting point
to derive the equation of motion for the the dynamics of the superfluid
at T = 0. In the BCS limit of a trapped Fermi superfluid, for example, such
a derivation has been carried out in detail by Baranov and Petrov\cite{Bara5}.
The analytic solutions of this T = 0  equation of motion describing
the collective modes of a BCS Fermi superfluid have been discussed by
Bruun and Clark\cite{Clark}.
\par
This paper is organized as follows: In Sec. \ref{sec2}, we briefly review\cite{Ohashi3,Ohashi5} our coupled fermion-boson (CFB) model in the presence of an isotropic harmonic trap. We calculate the generalized density correlation functions based on the HFB-GRPA formalism in Sec. \ref{sec4}. In Sec. \ref{sec6}, we summarize our previous results\cite{Ohashi5} for the single-particle excitation spectrum in a trapped superfluid Fermi gas, and then show in Sec. \ref{sec7} how the single-particle energy gap depresses the frequency of collective modes. Some of our results were briefly mentioned in Refs.\cite{Ohashi5,Ohashi-web}.
Appendix A gives a brief review of the results for the single-particle BCS excitations obtained in Ref. \cite{Ohashi5}. These are the input for the calculation of response functions in this paper. Appendix B proves that the renormalized Bose propagator for bound states has the same poles as the fermion density response function, as expected.
\par
\vskip5mm
\section{Coupled-Fermion Boson Model}
\label{sec2}
\vskip2mm
We consider a two-component superfluid Fermi gas with a Feshbach resonance trapped in a harmonic potential. To describe this, we consider a coupled fermion-boson (CFB) model given by\cite{Ohashi1,Ohashi2,Ohashi3,Milstein,Timmermans,Holland,Ohashi5,Ranninger}
\begin{eqnarray}
H
&=&
\sum_\sigma\int d{\bf r}
\Psi^\dagger_\sigma({\bf r})
\Bigl[
-{\nabla^2 \over 2m}+V_{\rm trap}^{\rm F}({\bf r})-\mu
\Bigr]
\Psi_\sigma({\bf r})
-U
\int d{\bf r}
\Psi^\dagger_\uparrow({\bf r})
\Psi^\dagger_\downarrow({\bf r})
\Psi_\downarrow({\bf r})
\Psi_\uparrow({\bf r})
\nonumber
\\
&+&
\int d{\bf r}
\Phi^\dagger({\bf r})
\Bigl[
-{\nabla^2 \over 2M}+2\nu+V_{\rm trap}^{\rm M}({\bf r})-2\mu
\Bigr]
\Phi({\bf r})
+
g_{\rm r}
\int d{\bf r}
\Bigl[
\Phi^\dagger({\bf r})\Psi_\downarrow({\bf r})\Psi_\uparrow({\bf r})+h.c.
\Bigr].
\nonumber
\\
\label{eq.2.1}
\end{eqnarray}
This model involves two field operators for Fermi atoms $\Psi_\sigma({\bf r})$ 
with pseudo-spin $\sigma=\uparrow,\downarrow$ and a Bose field operator $\Phi({\bf r})$ 
associated with the Feshbach resonance. 
The Feshbach resonance with strength $g_{\rm r}$ describes the conversion of two Fermi atoms into a molecular boson and the dissociation of a molecule into two Fermi atoms. This resonance is
controlled by the threshold energy $2\nu$ of molecular excitations, which is assumed to be adjustable. $U$ represents a non-resonant interaction, which we take to be attractive ($-U<0$). 
Since a molecule consists of two Fermi atoms, we take $M=2m$ and impose the conservation of the total number of atoms,
\begin{equation}
N=N_{\rm F}+2N_{\rm M}
=\sum_\sigma\int d{\bf r}
\langle\Psi^\dagger_\sigma({\bf r})\Psi_\sigma({\bf r})\rangle
+
2\int d{\bf r}\langle\Phi^\dagger({\bf r})\Phi({\bf r})\rangle.
\label{eq.2.2}
\end{equation}
This constraint has already been taken into account in Eq. (\ref{eq.2.1}) by
assuming that Bose chemical potential
$\mu_M\equiv2\mu$\cite{Ohashi1}, where $\mu$ is the Fermi chemical potential. 
In this paper, we assume that the harmonic trap potentials for Fermi atoms $V^{\rm F}_{\rm trap}(r)$ and Bose molecules $V^{\rm M}_{\rm trap}(r)$ are both isotropic, as
\begin{equation}
V_{\rm trap}^{\rm F}={1 \over 2}m\omega_0^2r^2,~~~
V_{\rm trap}^{\rm M}={1 \over 2}M\omega_{0M}^2r^2.
\label{eq.2.3}
\end{equation}
We also assume that atoms and molecules feel the same trap frequency ($\omega_{0M}=\omega_0$), which describes current experiments using Feshbach resonances.
\par
Since the CFB model in Eq. (\ref{eq.2.1}) involves Fermi atoms and 
Bose molecules, superfluidity is characterized by the BCS Cooper-pair potential
$\Delta({\bf r})\equiv\langle\Psi_\downarrow({\bf
  r})\Psi_\uparrow({\bf r})\rangle$ and the molecular BEC 
condensate $\phi_M({\bf r})\equiv\langle\Phi({\bf r})\rangle$. 
In equilibrium state, these order parameters are related to 
each other by\cite{Ohashi5}
\begin{equation}
{g_{\rm r} \over U}\Delta({\bf r})+
\Bigl[
-{\nabla^2 \over 2M}+V_{\rm trap}^{\rm M}(r)+2\nu-2\mu
\Bigr]\phi_M({\bf r})=0.
\label{eq.2.4}
\end{equation} 
This identity ensures that in the superfluid phase, $\Delta({\bf r})$ and $\phi_M({\bf r})$ are {\it both} finite in the same spatial region and vanish together.
\par
In discussing collective modes, 
it is convenient to divide Eq. (\ref{eq.2.1}) into the
Hartree-Fock-Bogoliubov (HFB) mean-field Hamiltonian ($\equiv H_{\rm HFB}$)
and a fluctuation part ($\equiv H_{\rm FL}$)\cite{OhashiT}. 
We introduce the Nambu representation for fermion field\cite{Sch}
\begin{equation}
{\hat \Psi}({\bf r})\equiv
\left(
\begin{array}{c}
\Psi_\uparrow({\bf r})\\
\Psi^\dagger_\downarrow({\bf r})
\end{array}
\right),
\label{eq.2.5}
\end{equation}
and its conjugate field ${\hat \Psi}^\dagger({\bf
  r})\equiv (\Psi^\dagger_\uparrow({\bf r}),\Psi_\downarrow({\bf
  r}))$. In this Nambu representation, the Fermi mean-field Hamiltonian $H_{\rm HFB}$ reduces to\cite{Ohashi5}
\begin{eqnarray}
H^F_{\rm HFB}
&=&
\int d{\bf r}
{\hat \Psi}^\dagger({\bf r})
\Biggl[\Bigl[
-{\nabla^2 \over 2m}-{U \over 2}n_{\rm F}({\bf r})+V_{\rm trap}^{\rm F}(r)-\mu
\Bigr]\tau_3
-{\tilde \Delta}({\bf r})\tau_1
\Biggr]
\Psi({\bf r}).
\label{eq.2.6}
\end{eqnarray}
Here $\tau_i$ ($i=1,2,3$) are the Pauli matrices acting in particle-hole space. $n_{\rm F}({\bf r})\equiv\sum_\sigma\langle\Psi_\sigma^\dagger({\bf r})\Psi_\sigma({\bf r})\rangle$ is the number density of Fermi atoms. We note that the fermion part $H_{\rm HFB}^{\rm F}$ has the same form as the standard BCS Hamiltonian\cite{Sch}, except that the Cooper-pair order parameter $\Delta({\bf r})$ is replaced with the {\it composite} order parameter, given by
\begin{equation}
{\tilde \Delta}({\bf r})\equiv\Delta({\bf r})-g_{\rm r}\phi_M({\bf
  r}).
\label{eq.2.7}
\end{equation}
In Eq. (\ref{eq.2.6}), the order parameter ${\tilde \Delta}({\bf r})$
is taken to be real and proportional to the $\tau_1$-component, without loss of generality. 
Because of the assumed spherical
symmetry of the trap, ${\tilde \Delta}({\bf r})$, $\Delta({\bf r})$, $\phi_M({\bf r})$, and $n_{\rm F}({\bf r})$ are all isotropic depending only on $r=|{\bf r}|$. 
\par
The molecular Bose Hamiltonian $H^M$ is given by the third term in Eq. (\ref{eq.2.1}). The part
\begin{eqnarray}
\delta H^M\equiv
\int d{\bf r}
\delta\Phi^\dagger({\bf r})
\Bigl[
-{\nabla^2 \over 2M}+2\nu+V_{\rm trap}^{\rm M}(r)-2\mu
\Bigr]\delta\Phi({\bf r})
\label{eq.2.6b}
\end{eqnarray}
describes fluctuations from the Bose condensate, where the molecular quantum field operator is defined as $\delta\Phi({\bf r})\equiv \Phi({\bf r})-\phi_M({\bf r})$.
\par
Fluctuation effects around the mean-field theory described by $H_{\rm HFB}^F$ can be expressed as 
interactions between fluctuations in the density-channel 
(particle-hole channel) and in the Cooper-channel (particle-particle channel)\cite{OhashiT}. It is convenient to introduce the {\it generalized density operators}, defined by\cite{note2}
\begin{equation}
{\hat \rho}_j({\bf r})={\hat \Psi}^\dagger({\bf r})\tau_j{\hat
  \Psi}({\bf r})~~~~(j=1,2,3).
\label{eq.2.8}
\end{equation}
Since we take $\Delta({\bf r})$ proportional to the $\tau_1$-component,
${\hat \rho}_1({\bf r})$ and ${\hat \rho}_2({\bf r})$ represent 
amplitude and phase fluctuations in the Cooper-channel, respectively. 
${\hat \rho}_3({\bf r})$ is the ordinary density
operator, expressed in the Nambu representation. Using ${\hat \rho}_j({\bf r})$,
we can write the non-resonant interaction $U$ as\cite{note3}
\begin{equation}
H_{\rm FL}^{(j)}=-{U \over 4}\int d{\bf r}{\hat \rho}_j({\bf r}){\hat
  \rho}_j({\bf r})~~~~(j=1,2,3).
\label{eq.2.9}
\end{equation}
Here $H_{\rm FL}^{(1)}$ and $H_{\rm FL}^{(2)}$ involve the the Cooper-channel, while
$H_{\rm FL}^{(3)}$ involves the density-channel. The Feshbach
resonance or the last term in Eq. (\ref{eq.2.1}) can be expressed as an interaction between fluctuations
in the molecular field $\delta\Phi({\bf r})$ and the Cooper-pair fluctuations ${\hat \rho}_1({\bf r})$ and ${\hat \rho}_2({\bf r})$, namely
\begin{eqnarray}
H_{\rm FL}^{F.R.}={g_{\rm r} \over 2}
\int d{\bf r}
\Bigl[
\delta\Phi^\dagger({\bf r}){\hat \rho}_-({\bf r})+\delta\Phi({\bf r}){\hat
  \rho}_+({\bf r})
\Bigr],
\label{eq.2.10}
\end{eqnarray}
where we define ${\hat \rho}_\pm({\bf r})\equiv{\hat \rho}_1({\bf r})\pm i{\hat \rho}_2({\bf r})$. Eq. (\ref{eq.2.10}) shows that the molecular fluctuations $\delta\Phi({\bf r})$ couple with the fluctuations ${\hat \rho}_1({\bf r})$ and ${\hat \rho}_2({\bf r})$ in the Cooper-channel, but do not couple with ordinary density fluctuations described by ${\hat \rho}_3({\bf r})$. 
\par
The fermion mean-field Hamiltonian $H^{\rm F}_{\rm HFB}$ in Eq. (\ref{eq.2.6}) can be diagonalized, reducing the problem to solving the Bogoliubov-de Gennes (BdG) coupled equations. In Ref. \cite{Ohashi5}, we have self-consistently solved these BdG equations and used the results to calculate the static properties as well as single-particle excitation spectrum in the BCS-BEC crossover, at $T=0$. Since the single-particle excitations are the ingredients in our calculations of the density correlation function, we summarize some of the discussion about BdG equations in Appendix A. For more details, see also the discussion in Ref. \cite{Ohashi5}.
\par
\vskip3mm
\section{Generalized density correlation functions}
\label{sec4}
\vskip2mm
Collective modes corresponds to the poles in the density correlation function. In the
superfluid phase, density fluctuations couple with particle-particle or Cooper-pair fluctuations\cite{OhashiT}, so that one needs to take their coupling into account. 
For this purpose, it is convenient to introduce the {\it generalized
density correlation functions}, defined by
\begin{equation}
\Pi_{ij}({\bf r},{\bf r}',\omega)=-i\int_0^\infty
dt e^{i\omega t}
\langle
[{\hat \rho}_i({\bf r},t),{\hat \rho}_j({\bf r}',0)]
\rangle.
\label{eq.2.35}
\end{equation}
$\Pi_{33}$ is the usual density correlation
function. $\Pi_{11}$ and $\Pi_{22}$ are the amplitude and phase
correlation functions, respectively. The off-diagonal components $\Pi_{i\ne
  j}$ describe couplings between different kinds of fluctuations. For
example, $\Pi_{23}$ is the phase-density coupling due to the
Josephson effect, as discussed in Refs. \cite{Ohashi3,OhashiT}.
\par
We calculate the generalized density correlation functions $\Pi_{ij}$
in a trap, based on the HFB-GRPA formalism in terms of the 
interactions $H_{\rm FL}^{(j)}$ 
in Eq. (\ref{eq.2.9}) and $H_{\rm FL}^{F.R.}$ in Eq. (\ref{eq.2.10}).
This is simply an extension of our previous work for a uniform superfluid 
Fermi gas\cite{Ohashi3} to include an isotropic harmonic trap.
In contrast to Ref. \cite{Ohashi3}, this paper considers the case $T=0$\cite{notegg}.
We refer to this earlier paper for a more detailed discussion of the physics.
The $3\times 3$ matrix correlation function ${\hat \Pi}\equiv\{\Pi_{ij}\}$ $(i,j=1,2,3)$  defined in Eq. (\ref{eq.2.35}) satisfies the GRPA equation in real space, namely
\begin{eqnarray}
{\hat \Pi}({\bf r},{\bf r}',\omega)
=
{\hat \Pi}^0({\bf r},{\bf r}',\omega)
&-&
{U \over 2}
\int d{\bf r}''
{\hat \Pi}({\bf r},{\bf r}'',\omega)
{\hat \Pi}^0({\bf r}'',{\bf r}',\omega)
\nonumber
\\
&+&
{g_{\rm r}^2 \over 2}
\int d{\bf r}''\int d{\bf r}'''
{\hat \Pi}({\bf r},{\bf r}'',\omega)
{\hat B}({\bf r}'',{\bf r}''',\omega)
{\hat \Pi}^0({\bf r}'',{\bf r}',\omega).
\label{eq.2.36}
\end{eqnarray}
In a matrix notation for the dependence on position, the {\it formal solution} of 
Eq. (\ref{eq.2.36}) is given by
\begin{equation}
{\hat \Pi}(\omega)={\hat \Pi}^0(\omega)
\Bigl[
1
+
{1 \over 2}
\Bigl[
U-g_{\rm r}^2{\hat B}(\omega)
\Bigr]
{\hat \Pi}^0(\omega)
\Bigr]^{-1}.
\label{eq.2.37}
\end{equation}
In Eq. (\ref{eq.2.36}), the effects of a Feshbach resonance appear as the {\it non-local} and  {\it energy-dependent} effective interaction 
${\hat B}({\bf r}'',{\bf r}''',\omega)$.
\par
${\hat B}$ has only the matrix elements $B_{11}$, $B_{12}$, $B_{21}$, and
$B_{22}$, indicating that it does not involve the ordinary density ${\hat \rho}_3({\bf r})$.
This originates from the fact that the Feshbach
resonance is described by an 
interaction between fluctuations of the molecular
Bose field $\delta\Phi({\bf r})$
and Cooper-pair order parameter fluctuations [see Eq. (\ref{eq.2.10})]. These
non-zero matrix elements can be expressed as
\begin{eqnarray}
\left(
\begin{array}{cc}
B_{11}({\bf r},{\bf r}',\omega) &
B_{12}({\bf r},{\bf r}',\omega) \\
B_{21}({\bf r},{\bf r}',\omega) &
B_{22}({\bf r},{\bf r}',\omega) 
\end{array}
\right)
={\hat P}{\hat D}^0({\bf r},{\bf r}',\omega){\hat P}^\dagger,
\label{eq.2.39}
\end{eqnarray}
where the $2\times 2$ matrix ${\hat P}$ is defined as
\begin{eqnarray}
{\hat P}=
{1 \over \sqrt{2}}
\left(
\begin{array}{cc}
1 & 1 \\
i & -i
\end{array}
\right)
\label{eq.2.40}
\end{eqnarray}
and ${\hat D}^0({\bf r},{\bf r}',\omega)$ is the $2\times 2$-matrix molecular Bose
Green's function given by
\begin{eqnarray}
{\hat D}^0({\bf r},{\bf r}',\omega)
=
\sum_{nlm}
{f^M_{nlm}({\bf r})f^M_{nlm}({\bf r}') \over
  \omega_+\tau_3-(E_{nl}^M+2\nu-2\mu)}
\equiv
\sum_{lm}Y_{lm}({\hat \theta}){\hat D}^{0l}(r,r',\omega)
Y_{lm}^*({\hat \theta}'),
\label{eq.2.41}
\end{eqnarray}
where the $l$-th component has the explicit form,
\begin{equation}
{\hat D}^{0l}(r,r',\omega)=\sum_{n}
{u_{nl}^M(r)u_{nl}^M(r') \over \omega_+\tau_3-(E_{nl}^M+2\nu-2\mu)}.
\label{eq.2.42}
\end{equation}
\par
In Eq. (\ref{eq.2.36}), ${\hat \Pi}^0({\bf r},{\bf
  r}',\omega)=\{\Pi^0_{ij}\}$ is the zeroth order $3\times 3$ matrix density
correlation function, obtained from the analytic continuation of the
corresponding two-particle thermal Green's function,
\begin{equation}
\Pi^0_{ij}({\bf r},{\bf r}',i\nu_m)=
{1 \over \beta}\sum_{\omega_m}
Tr\Bigl[
\tau_i{\hat G}({\bf r},{\bf r}',i\nu_m+i\omega_m)
\tau_j{\hat G}({\bf r}',{\bf r},i\omega_m)
\Bigr].
\label{eq.2.38}
\end{equation}
Here, $i\omega_m$ and $i\nu_m$ are the fermion and the boson Matsubara frequencies, respectively. The $2\times 2$ matrix single-particle Green's functions have the form,
\begin{equation}
{\hat G}({\bf r},{\bf r}',i\omega_m)=
-\int_0^\beta d\tau e^{i\omega_m\tau}
\Bigl\langle
T_\tau
\Bigl\{
{\hat \Psi}({\bf r},\tau){\hat \Psi}^\dagger({\bf r}',0)
\Bigr\}
\Bigr\rangle.
\label{eq.2.28}
\end{equation}
Using the eigenfunctions $f^F_{lmn}({\bf r})$ of the harmonic potential, 
we can write Eq. (\ref{eq.2.28})
\begin{equation}
{\hat G}({\bf r},{\bf r}',i\omega_m)=\sum_{lm}
Y_{lm}({\hat \theta}){\hat g}^l(r,r',i\omega_m)Y^*_{lm}({\hat \theta}'),
\label{eq.2.29}
\end{equation}
in terms of the fermion Green's function ${\hat g}^l(r,r',i\omega_m)$ 
for angular momentum $l$,
\begin{eqnarray}
{\hat g}^l(r,r',i\omega_m)
=
\sum_j
\Biggl[
{
\Lambda_{jl}(r)\Lambda_{jl}^\dagger(r')
\over 
i\omega_m-E^F_{jl}
}
+
{
{\bar \Lambda}_{jl}(r){\bar \Lambda}_{jl}^\dagger(r')
\over 
i\omega_m+E^F_{jl}
}
\Biggr].
\nonumber
\\
\label{eq.2.30}
\end{eqnarray}
Here we have introduced the wavefunctions related to BdG solutions
\begin{eqnarray}
\Lambda_{jl}(r)
\equiv
\sum_{n}
\left(
\begin{array}{c}
W^l_{n+1,j+1} \\
W^l_{{\bar N_l}+n,j+1}
\end{array}
\right)u^F_{nl}(r),~~~~{\bar \Lambda}_{jl}(r)=i\tau_2\Lambda_{jl}(r).
\label{eq.2.31}
\end{eqnarray}
\par
Substituting Eq. (\ref{eq.2.29}) into Eq. (\ref{eq.2.38}), one obtains\cite{Bruun1}
\begin{equation}
\Pi_{ij}^0({\bf r},{\bf r}',i\nu_m)=
\sum_{LM}Y_{LM}({\hat \theta})\Pi^{0L}_{ij}({\bf r},{\bf
  r}',i\nu_m)Y^*_{LM}({\hat \theta}').
\label{eq.2.43}
\end{equation}
Here, $\Pi^{0L}_{ij}({\bf r},{\bf r}',i\nu_m)$ describes the (generalized)
density fluctuations with angular momentum $L$,
\begin{equation}
\Pi^{0L}_{ij}({\bf r},{\bf r}',i\nu_m)=
\sum_{ll'}
{(2l+1)(2l'+1) \over 4\pi(2L+1)}
|\langle l0l'0|L0\rangle|^2
\Gamma_{ij}^{ll'}(r,r',i\nu_m),
\label{eq.2.44}
\end{equation}
where $\langle l0l'0|L0\rangle$ are the Clebsch-Gordan coefficients\cite{CG}. This 
leads to a selection rule in the summations with respect to $l$ and $l'$. 
The $r$-dependence of the correlation function is determined by $\Gamma_{ij}^{ll'}(r,r',i\nu_m)$, which has the form
\begin{eqnarray}
\Gamma_{ij}^{ll'}(r,r',i\nu_m)
&=&
{1 \over \beta}
\sum_{\omega_m}
Tr\Bigl[
\tau_i{\hat g}^l(r,r',i\nu_m+i\omega_m)
\tau_j{\hat g}^l(r',r,i\omega_m)
\Bigr]
\nonumber
\\
&=&
\sum_{nn'}
Tr
\Bigl[
\tau_i\Lambda_{nl}(r)\Lambda^\dagger_{nl}(r')
\tau_j{\bar \Lambda}_{n'l}(r'){\bar \Lambda}^\dagger_{n'l}(r)
\Bigr]
{1 \over i\nu_m-(E_{nl}^F+E_{n'l}^F)}
\nonumber
\\
&-&
\sum_{nn'}
Tr
\Bigl[
\tau_i{\bar \Lambda}_{nl}(r){\bar \Lambda}^\dagger_{nl}(r')
\tau_j\Lambda_{n'l}(r')\Lambda^\dagger_{n'l}(r)
\Bigr]
{1 \over i\nu_m+(E_{nl}^F+E_{n'l}^F)}.
\label{eq.2.45}
\end{eqnarray}
The correlation function $\Gamma_{ii}^{ll'}$ $(i=1,2,3)$, $\Gamma_{13}^{ll'}$ and $\Gamma_{31}^{ll'}$ can all be written in the form
\begin{eqnarray}
\Gamma_{ij}^{ll'}({\bf r},{\bf r}',i\nu_m)=-2
\sum_{nn'}Q_{ij}^{nn'}(r,r')
{E_{nl}^F+E_{n'l}^F \over \nu_m^2+(E_{nl}^F+E_{n'l}^F)^2},
\label{eq.2.46}
\end{eqnarray} 
where
\begin{eqnarray}
Q_{ij}^{nn'}(r,r')=
Tr
\Bigl[
\tau_2\tau_i
\Lambda_{nl}(r)\Lambda^\dagger_{nl}(r')
\tau_j\tau_2
\Lambda_{nl}(r')\Lambda^\dagger_{nl}(r)
\Bigr].
\label{eq.2.47}
\end{eqnarray}
The remaining components are given by
\begin{eqnarray}
\Gamma_{ij}^{ll'}({\bf r},{\bf r}',i\nu_m)=-2
\sum_{nn'}Q_{ij}^{nn'}(r,r')
{i\nu_m \over \nu_m^2+(E_{nl}^F+E_{n'l}^F)^2}.
\label{eq.2.48}
\end{eqnarray}
\par
After carrying out the usual analytic continuation, $i\nu_m\to\omega+i\delta$,
in Eq. (\ref{eq.2.43}), we substitute Eqs. (\ref{eq.2.41}) and (\ref{eq.2.43}) 
into Eq. (\ref{eq.2.36}). The result for ${\hat \Pi}({\bf r},{\bf
  r}',\omega)$ can be decomposed as
\begin{equation}
\Pi_{ij}({\bf r},{\bf r}',\omega_+)=
\sum_{LM}
Y_{LM}({\hat \theta})
{\hat \Pi}^{L}(r,r',\omega_+)
Y^*_{LM}({\hat \theta}'),
\label{eq.2.49}
\end{equation}
where the $L$-th component ${\hat \Pi}^L$ satisfies the GRPA equation
\begin{eqnarray}
{\hat \Pi}^L(r,r',\omega)
=
{\hat \Pi}^{0L}(r,r',\omega)
&-&
{U \over 2}
\int_0^\infty r''^2dr''
{\hat \Pi}^L(r,r'',\omega)
{\hat \Pi}^{0L}(r'',r',\omega)
\nonumber
\\
&+&
{g_{\rm r}^2 \over 2}
\int_0^\infty r''^2dr''\int_0^\infty r'''^2dr'''
{\hat \Pi}^L(r,r'',\omega)
{\hat B}^L(r'',r''',\omega)
{\hat \Pi}^{0L}(r'',r',\omega).
\nonumber
\\
\label{eq.2.50}
\end{eqnarray}
${\hat B}^L(r,r',\omega)$ is given by Eq. (\ref{eq.2.39}) but now with the
Bose propagator ${\hat D}^0(r,r',\omega)$ replaced with its
$L$-th component ${\hat
  D}^{0L}(r,r',\omega)$ defined in Eq. (\ref{eq.2.42}). In matrix notation for the position dependence, the
{\it formal} solution of Eq. (\ref{eq.2.50}) is given by
\begin{eqnarray}
{\tilde \Pi}^L(\omega)={\tilde \Pi}^{0L}(\omega)
\Bigl[
1
+
{1 \over 2}
\Bigl[
U-g_{\rm r}^2{\tilde B}^L(\omega)
\Bigr]
{\tilde \Pi}^{0L}(\omega)
\Bigr]^{-1},
\label{eq.2.51}
\end{eqnarray}
where 
${\tilde \Pi}^L(r,r',\omega)\equiv r{\hat \Pi}^L(r,r',\omega)r'$,
${\tilde \Pi}^{0L}(r,r',\omega)\equiv r{\hat
  \Pi}^{0L}(r,r',\omega)r'$, and ${\tilde B}^L(r,r',\omega)\equiv
r{\hat B}(r,r',\omega)r'$.
\par
The frequency of the collective mode in the $L$-th channel can be obtained from the pole of the response function ${\tilde \Pi}^L(r,r',\omega)$. The equation for the energy of the collective mode is given by the zero of the determinant
\begin{eqnarray}
det
\Bigg[
1
+
{1 \over 2}
\Bigl[
U-g_{\rm r}^2{\tilde B}^L(\omega)
\Bigr]
{\tilde \Pi}^{0L}(\omega)
\Biggr]=0.
\label{eq.3.3}
\end{eqnarray}
The dipole, quadrupole, and monopole modes are described by the $L=1$, $L=2$, and $L=0$ channels, respectively. All the collective modes accompanied by density fluctuations and/or fluctuations in the Cooper or particle-particle channel are given by the solutions of Eq. (\ref{eq.3.3}). For example, the gapless Goldstone mode at $\omega=0$ appears in the $L=0$ channel. We note that molecular boson Green's function has the same poles as those obtained from the solutions of Eq. (\ref{eq.3.3}), as proven in Appendix B. The physics behind this equivalence is that in the coupled fermion-boson model, fluctuations in the molecular field strongly couple into the fluctuations in the Cooper-channel through the Feshbach resonance. 
Thus, we need only calculate the density
correlation function for fermions to find the collective modes.
\par
The frequencies of the lowest energy collective modes are determined numerically by solving Eq. (\ref{eq.3.3}). The results will be discussed in Sec. \ref{sec7}. 
A weighted strength function for various density fluctuations is usefully defined as\cite{Bruun1}
\begin{eqnarray}
S_{33}(\omega,L)\equiv-
{1 \over \pi}
\int_0^\infty F_L(r) r^2dr\int_0^\infty
 F_L(r') r'^2 dr' 
Im
\Bigl[
\Pi^L_{33}(r,r',\omega)
\Bigr],
\label{eq.3.4}
\end{eqnarray}
where $F_1(r)=r$ for the dipole mode, $F_2(r)=r^2$ for the quadrupole mode, and $F_0(r)=r^2$ for the monopole mode\cite{Lundh}. 
These strength functions are evaluated in Sec. \ref{sec7}.
\vskip2mm

\vskip5mm
\section{Single particle excitations in a trapped superfluid Fermi gas}
\label{sec6}
\vskip2mm
In a recent paper\cite{Ohashi5}, we have numerically solved the BdG coupled equations together with Eq. (\ref{eq.2.26}), which determines the Fermi atom chemical potential $\mu$. We self-consistently determined the Cooper-pair order parameter $\Delta(r)$, the molecular condensate $\phi_M(r)$, the atom density $n_{\rm F}(r)$, and the chemical potential $\mu$. In the crossover region, we showed that the single-particle excitation gap $E_g\sim\omega_0$ is much smaller than the magnitude of the composite order parameter ${\tilde \Delta}(r=0)\sim\varepsilon_{\rm F}\gg\omega_0$ at the center of the trap. Since this is the key point in understanding the behavior of the collective mode frequencies in the BCS-BEC crossover, we briefly summarize the single-particle properties in a trapped Fermi superfluid in this section. The details are worked out in Ref. \cite{Ohashi5} and reviewed in Appendix A.
\par
In our numerical calculations, we take $N=10,912$, which is equivalent to filling up states up to the energy $E=31.5\hbar\omega_0$ ($\equiv\varepsilon_{\rm F}$) for a non-interacting Fermi gas in a trap. 
The Feshbach coupling is taken as $g_{\rm r}/\sqrt{R_{\rm F}^3}=0.06\omega_0$ (${\bar g}_{\rm r}\equiv g_{\rm r}\sqrt{N/R_{\rm F}^3}=0.2\varepsilon_{\rm F}$, where $R_{\rm F}\equiv\sqrt{2\varepsilon_{\rm F}/m\omega_0^2}$ is the Thomas-Fermi radius for a free Fermi gas in a trap). In order to see the effect of the single-particle excitation gap $E_g$ on collective modes, we consider the two cases for the nonresonant interaction $U$: $U/R_{\rm F}^3=0.001\omega_0$ (this corresponds to ${\bar U}\equiv UN/R_{\rm F}^3=0.35\varepsilon_{\rm F}$) and $0.0015\omega_0$ (${\bar U}=0.52\varepsilon_{\rm F}$). In the BCS region, the choice ${\bar U}=0.35\varepsilon_{\rm F}$ gives $E_g\sim\omega_0$ and the case ${\bar U}=0.52\varepsilon_{\rm F}$ gives $E_g\gg\omega_0$, where $\omega_0$ is the trap frequency. For the high energy cutoff, we take $\omega_c=161.5\hbar\omega_0$ ($\gg\varepsilon_{\rm F}$)\cite{notenum}. In order for the theory to be cutoff-free, we introduce the scattering length $a_s$ defined by\cite{Ohashi2,Ohashi5} ${4\pi a_s \over m}\equiv U_{\rm eff}/[1-U_{\rm eff}\sum_{0,\omega_c}{1 \over 2\varepsilon_{\bf p}}]$, where $\varepsilon_{\bf p}\equiv p^2/2m$ and $U_{\rm eff}\equiv U+g_{\rm r}^2/(2\nu+(3/2)\hbar\omega_0-2\mu)$ is the effective pairing interaction. For further discussion, see Ref. \cite{Ohashi5}.
\par
The single-particle density of states $N(\omega)$ is defined by
\begin{equation}
N(\omega\equiv-{1 \over \pi}
\int d{\bf r}
Im
\Bigl[
G_{11}({\bf r},{\bf r},i\omega_m\to\omega+i\delta)
\Bigr],
\label{eq.3.1}
\end{equation}
where $G_{11}$ is the (1,1)-component of the (analytic-continued) matrix single-particle Green's function defined in Eq. (\ref{eq.2.28}). 
From the spectrum shown in Fig. 1, one sees that a finite excitation energy gap occurs at $E_g\simeq 1.2\omega_0$. The threshold energy to break a Cooper-pair is given by $2E_{\rm g}$. In a {\it uniform} BCS superfluid, the Bogoliubov quasi-particle excitation spectrum is given by the well-known expression
\begin{equation}
E_{\bf p}=\sqrt{(\varepsilon_{\bf p}-\mu)^2+|{\tilde \Delta}|^2}.
\label{eq.3.1b}
\end{equation}
In the weak-coupling BCS regime, the chemical potential is close to the Fermi energy of a non-interacting Fermi gas, $\mu\simeq\varepsilon_{\rm F}>0$.  In this regime, the energy gap $E_g$ is equal to the order parameter, $E_g=|{\tilde \Delta}|$. This relation is no longer valid in the strong-coupling regime\cite{Leggett,Nozieres,Randeria}. When the chemical potential becomes {\it negative}, we find $E_g=\sqrt{\mu^2+|{\tilde \Delta}|^2}$. Moreover, the relation $E_g=|{\tilde \Delta}|$ is not valid even in the BCS regime when we deal with a trapped gas. Indeed, as shown in Fig. \ref{fig2}, the position-dependent composite order parameter ${\tilde \Delta}(r)=\Delta(r)-g_{\rm r}\phi_M(r)$ is nearly equal to $10\omega_0$ at the center of the trap, which is much larger than the value $E_g=1.2\omega_0$ shown in Fig. \ref{fig1}. Even at $r=0.5R_{\rm F}$, the order parameter is still much larger than $E_g$, with ${\tilde \Delta}(r=0.5R_{\rm F})\sim 5\omega_0\gg E_g$. 
\par
The difference between $E_g$ and ${\tilde \Delta}(r=0)$ is most remarkable in the crossover region. As shown in Fig. \ref{fig3}(a), the value of ${\tilde \Delta}(r=0)$ simply increases as one increases the strength of the (Feshbach-induced) pairing interaction. However, the energy gap $E_g$ is insensitive to this increase of the pairing interaction, always being of the order of the trap frequency $\omega_0$. 
When the chemical potential becomes negative in the BEC regime [see panel (b)], $E_g$ increases, approaching $E_g\sim |\mu|$\cite{Ohashi5}. In this BEC regime, $2E_g$ physically gives the threshold dissociation energy of a tightly bound molecule, i.e., $2E_g=|\mu_M|=2|\mu|$.
\par
The reason why the values of $E_g$ and ${\tilde \Delta}(r=0)$ are very different is because of the {\it Andreev bound states} appearing at the edge of the trap\cite{Ohashi5}. In a trapped superfluid, atoms feel the ``off-diagonal" pair potential ${\tilde \Delta}(r)$ (solid line in Fig. \ref{fig2}) as well as the ``diagonal" trap potential measured from the chemical potential\cite{noteH} 
\begin{equation}
V_{\rm eff}^{\rm trap}(r)\equiv (V_{\rm trap}^F(r)-\mu)\Theta(V_{\rm trap}^F(r)-\mu),
\label{eq.ap1}
\end{equation}
(dashed line in Fig. \ref{fig2}), where $\Theta(x)$ is the step function. Although ${\tilde \Delta}(r)$ and $V_{\rm eff}^{\rm trap}(r)$ appear in the diagonal and off-diagonal components, respectively, in the BdG coupled equations, they can be thought of roughly as giving rise to a combined potential well
\begin{equation}
V_{comb}(r)\equiv{\tilde \Delta}(r)+V_{\rm eff}^{\rm trap}(r).
\label{eq.ap2}
\end{equation}
As a result, the Andreev bound states (or in-gap states) appear well below ${\tilde \Delta}(r=0)$, being localized around the bottom of this combined potential well. The lowest bound state has energy $E_g$. This phenomenon is analogous to the appearance of surface bound states in superconductivity, originally discussed by de Gennes and Saint James\cite{Saint}. There is a large literature on these states.
\par
As in Ref. \cite{Ohashi5}, it is useful to also calculate the {\it local} density of states $N(\omega,r)$, defined by
\begin{equation}
N(\omega,r)=-{1 \over \pi}
Im
\Bigl[
G_{11}({\bf r},{\bf r},i\omega_m\to\omega+i\delta)
\Bigr].
\label{eq.3.2}
\end{equation}
Figure 4(a) exhibits the discrete in-gap (or Andreev) states at frequencies well below the order parameter at the center, $\omega\le{\tilde \Delta}(r=0)\simeq 10\omega_0$. These low energy states are localized at the edge of the trap $r\sim R_{\rm F}$. Such localized surface states are {\it not} given by a local density approximation (LDA) [see Fig. \ref{fig4}(b)]. In the LDA, the local density of states $N(\omega,r)$ is simply given by the uniform BCS density of states, with the order parameter and the chemical potential being replaced by the position-dependent ones, ${\tilde \Delta}(r)$ and $\mu(r)\equiv \mu-V_{\rm trap}^F(r)$, respectively. The ridge peak in the LDA density of states in Fig. 4(b) simply corresponds to ${\tilde \Delta}(r)$. 
\par
We note that the position of an in-gap state depends on the magnitude of the off-diagonal pair potential ${\tilde \Delta}(r)$ at the center of the trap. We have seen that the bound state wavefunction is well localized at the edge of the trap when ${\tilde \Delta}(r=0)\gg\omega_0$. On the other hand, the surface bound states penetrate deep inside the trap when ${\tilde \Delta}(r=0)$ is small. In Fig. \ref{fig5}(a) [where ${\tilde \Delta}(r=0)\simeq 1.86\omega_0$], the local density of states at the center of the trap ($r=0$) has a large intensity below $\omega={\tilde \Delta}(r=0)\simeq 1.86\omega_0$, originating from the Andreev bound state.
\par
In contrast, in the BEC regime [$(k_{\rm F}a_s)^{-1}<0$], where the chemical potential $\mu$ is negative, the low-lying bound states at the edge of the trap become less important for the single-particle excitation spectrum. As expected from Eq. (\ref{eq.3.1b}) for a uniform Fermi gas, the threshold of the excitation spectrum is then dominated by the chemical potential $|\mu|$. There are no Andreev bound states in the local density of states shown in Fig. \ref{fig5}(b). Instead, a very large excitation gap appears, due to the large dissociation energy ($\sim 2|\mu|$) of a molecule.
\vskip5mm
\section{Collective mode frequencies in a trapped Fermi superfluid}
\label{sec7}
\vskip3mm
\subsection{Quadrupole mode}
\vskip2mm
In the $T=0$ hydrodynamic theory appropriate to a pure superfluid, a collective mode of a superfluid Fermi gas in the BCS weak-coupling limit in an isotropic harmonic trap has the frequency\cite{Pethick,Bara5}
\begin{equation}
\omega=\omega_0\sqrt{L+{4 \over 3}n(2+L+n)},
\label{eq.3.5}
\end{equation}
where $L$ and $n$ are the angular momentum and radial quantum numbers of the 
collective mode, respectively. This expression predicts that the frequency of the quadrupole mode ($L=2$, $n=0$) is $\omega=\sqrt{2}\omega_0$. 
Figure \ref{fig6} shows the frequency of the quadrupole mode in the BCS-BEC crossover, as given by our calculations summarized in Sec. \ref{sec4}. 
The quadrupole mode frequency is suppressed so as to stay below the two-particle continuum threshold given by $2E_g$, where $E_g$ is the single-particle energy gap calculated in Sec. \ref{sec4}. 
In the BCS regime, our numerical results in Fig. 6(a) for the quadrupole mode agree with the prediction $\omega=\sqrt{2}\omega_0$. In this case (${\bar U}=0.52\varepsilon_{\rm F}$), the threshold energy of two-particle continuum $2E_g$ is greater than the collective mode frequency $\sqrt{2}\omega_0$. However, when $2E_g<\sqrt{2}\omega_0$, Fig. \ref{fig6}(b) shows a strong deviation from the predicted hydrodynamic frequency $\omega=\sqrt{2}\omega_0$ even in the BCS regime. We note that quadrupole mode in our calculation has no damping (due to decay into two quasiparticles) in the entire BCS-BEC crossover region, since the frequency always lies below the computed values of $2E_g$. 
\par
The suppression of the quadrupole mode frequency due to the two-particle continuum shown in Fig. \ref{fig6}(b) continues until we enter the BEC regime, where the chemical potential becomes large and negative and hence $2E_g$ also becomes large. In Ref. \cite{Ohashi5}, we showed that in the BCS-BEC crossover region, the combined potential well in Eq. (\ref{eq.ap2}) becomes wider in the crossover region. This leads to a decrease in the energy of the lowest Andreev level, which in turn determines the single-particle energy gap $E_g$. Reflecting this phenomenon, in Fig. \ref{fig6}(b), we note that the suppression of the quadrupole mode frequency is strongest around $(k_{\rm F}a_s)^{-1}\simeq 0.5$, where the value of $E_g$ is a minimum. 
\par
Since the single-particle excitation gap $E_g$ becomes very large $(\gg\omega_0)$ when $\mu<0$ (see Fig. \ref{fig3}), the effect of the two-particle continuum on the quadrupole mode is weak in the BEC regime. As a result, in Fig. \ref{fig6}(b), the frequency of the quadrupole mode shows a sharp increase as we enter this region. In the extreme BEC limit [$(k_Fa_s)^{-1}\to\infty$], our quadrupole mode frequency approaches the energy level of a non-interacting molecular Bose gas with $L=2$ and $n=0$, which is given by $\omega=(2n+L)\omega_0=2\omega_0$. Thus our present calculations do not give the correct quadrupole mode frequency in the BEC limit for an interacting Bose-condensed gas of molecules\cite{Stringari2}. We briefly explain the reason for this. 
\par
In our present approximation, the effective interaction between molecules ($\equiv V^M_{\rm eff}$) in the BEC regime is mediated by free Fermi atoms, as shown in Fig. \ref{fig7}. In a {\it uniform} superfluid Fermi gas, taking incident and scattered molecules to be at zero momentum, this process gives the following effective interaction
\begin{eqnarray}
V_{\rm eff}^M
=
{g_{\rm r}^4 \over \beta}\sum_{{\bf p},i\omega_m}
G^2_{11}({\bf p},i\omega_m)G^2_{11}(-{\bf p},-i\omega_m)
=
{g_{\rm r}^4N \over 4}{3\pi \over 32}{1 \over (\varepsilon_{\rm F}\nu)^{3/2}}.
\label{eq.eff}
\end{eqnarray}
It is interesting to note that this expression for the repulsive molecular interaction is precisely in agreement with that computed from the velocity of the Goldstone phonon in the BEC region of a {\it uniform} Fermi superfluid 
(see Eq. (4.20) of Ref. \cite{Ohashi3}). 
As we have discussed earlier, in a {\it trapped} Fermi superfluid, our present calculations are limited to a weak Feshbach resonance, where the coupling parameter ${\bar g}_{\rm r}=0.2\varepsilon_{\rm F}<\varepsilon_{\rm F}$, due to computational problems. In this case, 
by analogy to the uniform gas result in Eq. (\ref{eq.eff}), we expect $V_{\rm eff}^M~(\propto g_{\rm r}^4)$ to be very small. This explains why in our results in Fig. \ref{fig6}, the quadrupole mode frequency approaches the non-interacting Bose gas result $\omega=2\omega_0$ in the BEC region dominated by molecules (where $2E_g$ is large and $\mu<0$). 
\par
For a broad Feshbach resonance (${\bar g}_{\rm r}\gg\varepsilon_{\rm F}$), in contrast, one can expect a stronger effective interaction $V_{\rm eff}^M$ between molecules. 
In this case, in the BEC regime, we would expect that the collective modes would be those of trapped Bose superfluid at $T=0$\cite{Pethick,Pita2,Stringari2}, namely
\begin{equation}
\omega=\omega_0\sqrt{L+3n+2nL+2n^2}.
\label{eq.3.6}
\end{equation}
This $T=0$ hydrodynamic superfluid theory gives $\omega=\sqrt{2}\omega_0$ for the quadrupole mode in the molecular BEC region, the same as in the BCS limit (in contrast to a uniform gas).
\par
Since the threshold energy $2E_g$ becomes small near the superfluid transition temperature $T_{\rm c}$\cite{Ohashi1,Ohashi2}, the suppression of the quadrupole mode frequency by the two-particle continuum will become more important at finite temperatures. Even in the case shown in Fig. \ref{fig6}(a), the quadrupole mode frequency will deviate from the simple prediction $\omega=\sqrt{2}\omega_0$ in the BCS and the crossover region, as soon as the threshold energy of the two-particle continuum $2E_g(T)$ is greater than $\sqrt{2}\omega_0$. Thus, observing the suppression effect on the quadrupole mode frequency by the two-particle continuum might give useful information about the temperature dependent single-particle energy gap $E_g(T)$. 
\par
Figures \ref{fig8} is a plot of the strength function $S_{33}(\omega,L=2)$ defined in Eq. (\ref{eq.3.4}) for the parameters used in Fig. 6(b). In each panel, the lowest energy peak is the quadrupole mode. In panels (a)-(c), since the threshold energy of the two-particle continuum (shown by the arrows) is close to the frequency of the quadrupole mode, we can also see fine structure at frequencies just above the quadrupole mode, associated with the two-particle excitation spectrum. On the other hand, in the case ${\bar U}=0.52\varepsilon_{\rm F}$, the threshold energy $2E_g$ of the two-particle excitation spectrum is much higher than the frequency of the quadrupole mode, as shown in Fig. 6(a). As a result, only a peak associated with the quadrupole mode appears in the low-energy spectrum of the strength function, as shown in Fig. 9. There is no fine structure associated with the two-particle continuum, of the kind shown in Fig. \ref{fig8}.
\par
\vskip3mm
\subsection{Monopole mode}
\vskip2mm
Figure 10 shows the frequency of the monopole mode in the BCS-BEC crossover region. This collective mode is also suppressed below the two-particle continuum at $2E_g$. (Note that the BCS superfluid hydrodynamic prediction in Eq. (\ref{eq.3.5}) gives $\omega=2\omega_0$ for $L=0$ and $n=1$). Thus, as in case of the quadrupole mode, the monopole mode is not damped at $T=0$. 
In the BEC regime ($\mu<0$), the effect of two-particle continuum on the monopole mode frequency is absent because $2E_g\gg\omega_0$. In the BEC limit, our calculated monopole mode frequency approaches $\omega=2\omega_0$, the excitation frequency for a non-interacting Bose gas with $L=0$ and $n=1$. 
The quadrupole mode has the identical feature, for reasons discussed above. When the effective interaction between molecules is significant, as expected in the case of a broad Feshbach resonance, the monopole mode frequency would approach $\omega=\sqrt{5}\omega_0=2.24\omega_0$ in an improved theory\cite{Ohashi7}. 
\par
The monopole mode appears as the lowest energy peak in the spectrum of the strength function $S_{33}(\omega,L=0)$ for the $L=0$ channel, as shown in Figs. \ref{fig11} and \ref{fig12}. As expected, we can see the spectral weight coming from the two-particle excitations near the monopole peak shown in panels (a)-(c) of Fig. \ref{fig11}. 
\par
The situation is quite different in the dipole (Kohn) mode\cite{Ohashi4,Kohn}. This collective mode is the rigid oscillation of the center of motion, so that the interaction between atoms, dependent only on the relative coordinate of two atoms, has no effect on the frequency of this mode. As a result, the dipole mode {\it always} appears at the trap frequency $\omega_0$, irrespective of the value of the two-particle continuum $2E_g$.
\par
\subsection{Comparison to experiment}
\par
In concluding this section, we would like to briefly discuss the possible relevance of our prediction to current experimental results on collective modes in trapped Fermi superfluids.
The radial breathing modes have been studied experimentally in a $^6$Li superfluid in a cigar shaped trap\cite{Kinast,Bar2}. The observed modes have small damping ($\Gamma$) in the BEC and crossover region, but $\Gamma$ suddenly increased precisely where frequency of the mode suddenly increased\cite{Bar2}. As the unitarity limit $[(p_{\rm F}a_s)^{-1}=0]$ is approached from the BEC side, the radial breathing mode frequency always decreases, in contrast to the prediction by the $T=0$ superfluid hydrodynamic theory\cite{Stringari}. At the unitarity limit, the frequency of the radial breathing mode in Ref. \cite{Kinast} agrees with the ``universal" value ($\omega_{unitarity}$) predicted by superfluid hydrodynamic theory. However, in Ref. \cite{Bar2}, the observed radial breathing mode frequency was found to be smaller than $\omega_{unitarity}$.
Although our calculations are for an isotropic trap, the suppression effect we obtain for the collective mode frequencies by the two-particle excitation threshold should also arise in anisotropic traps. 
This is ignored in the $T=0$ equation of motion for a pure superfluid used to obtain the frequency $\omega_{unitarity}$. 
The decrease of the mode frequency which we predict as we approach the unitarity limit from the BEC side is thus qualitatively consistent with the experimental results in superfluid $^6$Li\cite{Kinast,Bar2}. 
\par
The small damping of the breathing modes in superfluid $^6$Li\cite{Bar2} may also be related to the ``suppression effect" we have discussed in this paper, i.e., collective modes always stay below the two-particle excitation threshold and thus are stable with no damping at $T=0$. On the other hand, our results show that the frequencies of the collective modes smoothly change as we pass through the crossover region, and does not predict the observed sudden increase of the radial breathing mode frequency in the BCS regime of superfluid $^6$Li\cite{Bar2}. A breakdown of the superfluid phase or some other cause would seem to be the source of this sudden increase in damping, but this requires further study. 
\vskip5mm
\section{Summary}
\vskip2mm
In this paper, using a collisionless HFB-GRPA type theory at $T=0$, we have investigated the collective modes in a trapped superfluid Fermi gas with a Feshbach resonance. 
In essence, we have extended our previous work\cite{Ohashi3} on the collective modes in a {\it uniform} Fermi superfluid to include an isotropic harmonic trap. 
Our present calculation of the collective modes makes crucial use of our recent work\cite{Ohashi5} on the single-particle excitations of a trapped Fermi superfluid, given by the solutions of the Bogoliubov-de Gennes coupled equations. Using these eigenstates, we have calculated the generalized density correlation function in the HFB-GRPA formalism, which has been shown\cite{Ohashi4} to be consistent with both Goldstone's theorem and Kohn's theorem in the crossover region of a trapped Fermi superfluid. 
\par
One of the major results of the present paper is that we have shown in detail how quadrupole and monopole mode frequencies are suppressed by the threshold energy of the two-particle continuum (twice the single-particle excitation gap $E_g$). When $2E_g$ is smaller than the collective mode energy ($\equiv\omega_{\rm h.d.}$) predicted by a pure superfluid (or hydrodynamic) theory, the mode frequency can deviate significantly from $\omega_{h.d.}$. As we showed in Ref. \cite{Ohashi5} (see also Fig. \ref{fig3} of this paper), the single-particle excitation gap $E_g$ is of the order of the trap frequency $\omega_0$ in a trapped Fermi superfluid in the crossover region. This is much smaller than the magnitude of the composite order parameter ${\tilde \Delta}(r=0)$ at the center of the trap, which is comparable to the Fermi energy $\varepsilon_{\rm F}$. This is because $E_g$ is determined by the lowest energy Andreev bound state localized at the edge of the trap\cite{Ohashi5,Baranov}. The suppression of the collective mode by the two-particle continuum should be observable, and would be even more 
important at temperatures close to $T_{\rm c}$.
\par
We have emphasized that the use of the $T=0$ equation of motion of a pure superfluid\cite{Pita,Stringari2} (which assumes that the density of superfluid equals the total density) precludes any inclusion of effects of the two-particle continuum. Thus, the work presented here extends previous work. 
We also note that our work is built on the calculation (see Ref. \cite{Ohashi5} and Appendix A) of the correct BdG BCS excitations of a trapped Fermi superfluid. This allow us to highlight the feature that the single-particle energy gap $E_g$ arises from low energy BCS excitations, localized near the edge of the trap. These states, by definition, are left out any theory based on the local density approximation (LDA).
\par
In the BEC regime, where the Fermi chemical potential $\mu$ is negative, the two-particle excitation spectrum exhibits a large gap, with $2E_g\simeq 2|\mu|\gg\omega_0$. In this limit, the low-energy collective modes with $\omega\sim\omega_0$ are no longer much effected by the higher energy two-particle continuum.
\par
Our results in this paper may be useful as a way to obtain useful information about the single-particle excitation gap energy $E_g\sim \omega_0$ in the crossover region. Recently, the single-particle Fermi excitation spectrum was studied using the rf-tunneling spectroscopy in $^6$Li\cite{Chin}. In this experiment, the energy gap is evaluated from an emergent peak in the spectrum as the gas becomes superfluid. This observed gap actually only gives information about the magnitude of the composite order parameter ${\tilde \Delta}(r=0)$ in the center of the trap, which is of the order of the Fermi energy $\varepsilon_{\rm F} (\gg\omega_0)$ in the crossover region. In principle, such rf-tunneling experiments should be able to observe the true excitation gap at much smaller energies $E_g\sim\omega_0$, but this would require very good energy resolution. Observing the suppression of a collective mode frequency discussed in this paper due to the presence of the two-particle continuum may be an alternative way to measure the magnitude of $E_g$. Although we have assumed an isotropic trap in this paper, the same suppression will also occur in the anisotropic trap used in recent experiments. 
\par
In this paper, we have only considered the case of $T=0$. At finite temperatures, Bogoliubov quasi-particles are excited thermally, which will give rise to Landau damping of the collective modes. In addition, fluctuation effects from the Cooper-channel\cite{Ohashi1,Ohashi2,Ohashi3,Nozieres,Randeria} become crucial, especially near $T_{\rm c}$. In future work, we hope to extend the present results at finite temperatures.
\par
\vskip2mm
\acknowledgments
Both authors acknowledge support through a research grant from NSERC of Canada. Y. O. was also financially supported by a Grant-in-Aid for Scientific research from the Ministry of Education, Culture, Sports, Science and Technology of Japan, as well as by a University of Tsukuba Research Project. 
\par 

\appendix
\vskip3mm
\section{Single-particle excitations in the BCS-BEC crossover region}
\label{sec3}
\vskip3mm
In this appendix, we review some key results in Ref. \cite{Ohashi5} for the single-particle excitations. We expand the fermion field operator $\Psi_\sigma({\bf r})$ in the eigenfunctions $f^F_{nlm}({\bf r})\equiv u^F_{nl}(r)Y_{lm}({\hat \theta})$ of the harmonic potential $V_{\rm trap}^F(r)$, as
\begin{equation}
{\hat \Psi}_\sigma({\bf r})=\sum_{nlm}f^F_{nlm}({\bf r})c_{nlm\sigma}.
\label{eq.2.10b}
\end{equation}
Here $Y_{lm}({\hat \theta})$ is a spherical harmonic, and the radial wavefunction has the form
\begin{equation}
u^F_{nl}(r)=
\sqrt{2}(m\omega_0)^{3/4}
\sqrt{n! \over (n+l+1/2)!}
e^{-{{\bar r}^2 \over 2}}
{\bar r}^l
L_n^{l+1/2}({\bar r}^2)~~~({\bar r}\equiv\sqrt{m\omega_0}r),
\label{eq.2.11}
\end{equation}
where $L_n^{l+1/2}({\bar r}^2)$ is the Laguerre polynomial. Substituting Eq. (\ref{eq.2.10b}) into $H_{\rm HFB}^{\rm F}$ in Eq. (\ref{eq.2.6}), we obtain $H_{\rm HFB}^{\rm F}=\sum_{lm}H_{\rm HFB}^{lm}$., where the $(l,m)$-component $H_{\rm HFB}^{lm}$ is given by
\begin{eqnarray}
H_{\rm HFB}^{lm}
=\sum_{n,\sigma}\xi_{nl}^{\rm F}c^\dagger_{nlm\sigma}c_{nlm\sigma}
&-&
{U \over 2}
\sum_{nn',\sigma}J_{nn'}^l
c^\dagger_{nlm\sigma}c_{n'lm\sigma}
\nonumber
\\
&-&
\sum_{nn'}F_{nn'}^l
\Bigl[
c^\dagger_{nlm\uparrow}c^\dagger_{n'l-m\downarrow}+h.c.
\Bigr].
\label{eq.2.12}
\end{eqnarray}
Here $\xi^{\rm F}_{nl}=\hbar\omega_0(2n+l+3/2)-\mu$ is the eigenenergy of $f^F_{nlm}({\bf r})$. $F_{nn'}^l$ and $J_{nn'}^l$ represent the scattering matrix elements involving the composite pair potential ${\tilde \Delta}(r)$ and the Hartree potential $-{U \over 2}n_{\rm F}(r)$, respectively. They are defined by
\begin{equation}
F_{nn'}^l=\int_0^\infty
 r^2dr u^F_{nl}(r){\tilde \Delta}(r)u^F_{n'l}(r),
\label{eq.2.13}
\end{equation}
\begin{equation}
J_{nn'}^l=\int_0^\infty
 r^2dr u^F_{nl}(r)n_{\rm F}(r)u^F_{n'l}(r).
\label{eq.2.14}
\end{equation}
The matrix element $F_{nn'}^l$ in Eq. (\ref{eq.2.13}) includes both intrashell pairing $(n=n')$ and intershell pairing ($n\ne n'$). 
\par
The quadratic Hamiltonian in Eq. (\ref{eq.2.12}) can be diagonalized by the unitary transformation
\begin{eqnarray}
\left(
\begin{array}{c}
c_{0lm\uparrow} \\
\ldots \\
c_{N_llm\uparrow} \\
c^\dagger_{0l,-m\downarrow} \\
\ldots \\
c^\dagger_{N_ll,-m\downarrow} \\
\end{array}
\right)=
{\hat W}^l
\left(
\begin{array}{c}
\gamma_{0lm\uparrow} \\
\ldots \\
\gamma_{N_llm\uparrow} \\
\gamma^\dagger_{0lm\downarrow} \\
\ldots \\
\gamma^\dagger_{N_llm\downarrow} \\
\end{array}
\right),
\label{eq.2.22}
\end{eqnarray}
where the $2(N_l+1)\times 2(N_l+1)$-orthogonal matrix ${\hat W}^l$ is taken so that ${\hat W}^{l\dagger} H_{\rm HFB}^{lm}{\hat W}^l$ be diagonal. This requirement leads to the BdG coupled equations for the wavefunction $W^l_{i,n}$, where $i,n$ can go from $0$ to $2(N_l+1)$ [see Eq. (2.25) of Ref.\cite{Ohashi5}].
In Eq. (\ref{eq.2.22}), $\gamma_{jlm\sigma}$ represents the annihilation operator of a Bogoliubov quasiparticle excitation, with eigenenergy $E_{nl}^F$. The resulting diagonalized $H_{\rm HFB}^{lm}$ has the form (constant terms are dropped)
\begin{equation}
H_{\rm HFB}^{lm}=\sum_{j=0}^{N_l}E^F_{nl}\gamma_{jlm\sigma}^\dagger\gamma_{jlm\sigma}~~~(E^F_{nl}>0).
\label{eq.2.21b}
\end{equation}
As usual in the BCS theory, one needs a cutoff $\omega_c\equiv\hbar\omega_0(N_c+3/2)$ to avoid ultraviolet divergences. The {\it maximal} radial quantum number $N_l$ in Eq. (\ref{eq.2.21b}) is given by the largest integer bounded by $(N_c-l)/2$. Using the solutions of the BdG equations, the Cooper-pair order parameter $\Delta(r)$ and the Fermi atom density $n_{\rm F}(r)$ are given by the solutions $W^l_{i,n}$ of the BdG equations\cite{Ohashi5}
\begin{eqnarray}
\Delta(r)=U
{\sum_{nn'l}}'{2l+1 \over 4\pi}
u_{nl}(r)u_{n'l}(r)
\sum_{j=0}^{N_l}
W_{{\bar N}_l+n,{\bar N}_l+j}^l
W_{n'+1,{\bar N}_l+j}^l.
\label{eq.2.23}
\end{eqnarray}
\begin{eqnarray}
n_{\rm F}(r)=
\sum_{nn'l}{2l+1 \over 4\pi}
u_{nl}(r)u_{n'l}(r)
\sum_{j=0}^{N_l}
\Bigl[
W_{n+1,{\bar N}_l+j}^l
W_{n'+1,{\bar N}_l+j}^l
+
W_{{\bar N}_l+n,j+1}^l
W_{{\bar N}_l+n',j+1}^l
\Bigr].
\label{eq.2.24}
\end{eqnarray}
Here ${\bar N}_l=N_l+2$ and the prime in ${\sum}'$ refers to the cutoff $\omega_c$. 
\par
Since the molecular condensate component $\phi_M({\bf r})$ is isotropic, it can be expanded in terms of the radial component $u^M_{nl}(r)$\cite{noteB} of the wavefunction $f^M_{nlm}({\bf r})\equiv u^M_{nl}(r)Y_{lm}({\hat \theta})$ for a Bose molecule in the harmonic potential $V_{\rm trap}^M(r)$. We obtain
\begin{eqnarray}
\phi_M(r)
&=&
{1 \over \sqrt{4\pi}}\sum_n\alpha_n u_{n0}^M(r)
\nonumber
\\
&=&
-{g_{\rm r} \over U}
\sum_n
{u_{n0}^M(r) \over E_{n0}^M+2\nu-2\mu}
\int_0^\infty dr'{r'}^2
\Delta(r'),
\label{eq.2.17}
\end{eqnarray}
where $E^M_{nl}=\hbar\omega_0(2n+l+3/2)$ is the energy of the molecular state $f^M_{nlm}({\bf r})$. The last expression in Eq. (\ref{eq.2.17}) is obtained by using 
the identity in Eq. (\ref{eq.2.4}). 
\par
The boson mean-field Hamiltonian $\delta H^M$ in Eq. (\ref{eq.2.6b}) can be easily diagonalized, reducing to
\begin{equation}
\delta H^M=\sum_{nlm}(E_{nl}^M+2\nu-2\mu)b_{nlm}^\dagger b_{nlm},
\label{eq.2.20}
\end{equation}
where $b_{nlm}$ is the annihilation operator of the molecular eigenstate $f^M_{nlm}({\bf r})$. 
\par
To describe the BCS-BEC crossover region at $T=0$, we extend the strong-coupling theory developed by Leggett\cite{Leggett}, to the case of superfluidity in a Fermi gas trapped in a harmonic potential with a Feshbach resonance. The key is that one needs to take into account the deviation of the chemical potential $\mu$ from the Fermi energy due to the formation of a bound state Bose condensate\cite{Leggett}. This effect is incorporated into the theory at $T=0$ by solving the BdG equations together with the equation for the number of Fermi atoms. In the presence of a Feshbach resonance, a fraction of these atoms form molecular bosons. At $T=0$, our discussion assumes all these molecules are Bose condensed, so that the total density of atoms is given by 
\begin{equation}
n(r)=2|\phi_M(r)|^2+n_{\rm F}(r)\equiv 2n_{\rm M}(r)+n_{\rm F}(r).
\label{eq.2.25}
\end{equation}
The equation of the state $\mu(N)$ giving $\mu$ as a function of the total number of atoms to determine $\mu$ is obtained by
integrating Eq. (\ref{eq.2.25}) over ${\bf r}$. The result is
\begin{equation}
N=2\sum_n\alpha_n^2
+\sum_{nn'l}(2l+1)
\Bigl[
|W_{n+1,{\bar N}_l+n'}^l|^2+
|W_{{\bar N}_l+n,n'+1}^l|^2
\Bigr].
\label{eq.2.26}
\end{equation}
The self-consistent solution $W^l_{i,n}$ and (\ref{eq.2.26}) are the basis of our mean-field crossover theory at $T=0$\cite{Ohashi5}. The resulting BCS excitations are used to calculate the response functions in the text of this paper.
\par
\vskip5mm
\section{Molecular Bose propagator}
\label{sec5}
\vskip2mm
The renormalized $2\times 2$ matrix Bose Green's function describing molecules, 
which includes the 
Feshbach resonance effect within the GRPA, is given by\cite{Ohashi3} (we
use a matrix notation for ${\bf r}$ and ${\bf r}'$)
\begin{eqnarray}
{\hat D}(\omega)=
\Bigl[
1-{\hat \Sigma}(\omega){\hat D}^0(\omega)
\Bigr]^{-1}
{\hat D}^0(\omega).
\label{eq.2.52}
\end{eqnarray}
The explicit form for ${\hat D}^0(\omega)$ is given in Eq. (\ref{eq.2.41}).
The self-energy of the molecules ${\hat \Sigma}(\omega)$ is due to the 
fluctuations associated with the particle-particle fluctuations in the Fermi gas,
\begin{equation}
{\hat \Sigma}(\omega)={g_{\rm r}^2 \over 2}
{\hat P}^\dagger
{\hat \eta}
\Biggl[
{\hat \Pi}^0(\omega)
\Bigl[
1+{U \over 2}{\hat \Pi}^0(\omega)
\Bigr]^{-1}
\Biggr]
{\hat P}.
\label{eq.2.53}
\end{equation}
Here we have introduced a projection operator ${\hat \eta}[{\hat A}]$ which 
extracts the
(1,1), (1,2), (2,1), and (2,2) components from a $3\times 3$-matrix
${\hat A}$. The $2\times 2$ matrix ${\hat P}$ is defined in Eq. (\ref{eq.2.40}).
\par
Excitations from the molecular condensate are determined from the poles
of the renormalized Bose Green's function ${\hat D}(\omega)$ in Eq. (\ref{eq.2.52}). 
This has been discussed in Ref.\cite{Ohashi3} for the case of a uniform superfluid Fermi gas.
The excitation energy of the dressed molecule 
is obtained from the zeros of the determinant
\begin{eqnarray}
0
&=&
det
\Biggl[
1-{g_{\rm r}^2 \over 2}{\hat D}^0(\omega)
{\hat P}^\dagger
{\hat \eta}
\Biggl[
{\hat \Pi}^0(\omega)
\Bigl[
1+{U \over 2}{\hat \Pi}^0(\omega)
\Bigr]^{-1}
\Biggr]
{\hat P}
\Biggr]
\nonumber
\\
&=&
det
\Biggl[
1-{g_{\rm r}^2 \over 2}{\hat B}(\omega)
{\hat \Pi}^0(\omega)
\Bigl[
1+{U \over 2}{\hat \Pi}^0(\omega)
\Bigr]^{-1}
\Biggr]
\nonumber
\\
&=&
det
\Biggl[
\Bigl[
1+{U \over 2}{\hat \Pi}^0(\omega)
-{g_{\rm r}^2 \over 2}{\hat B}(\omega)
{\hat \Pi}^0(\omega)
\Bigr]
\Bigl[
1+{U \over 2}{\hat \Pi}^0(\omega)
\Bigr]^{-1}
\Biggr]
\nonumber
\\
&=&
{det
\Biggl[
1+{1 \over 2}
\Bigl[
U-g_{\rm r}^2{\hat B}(\omega)
\Bigr]
{\hat \Pi}^0(\omega)
\Biggr]
\over
det\Bigl[1+{U \over 2}{\hat \Pi}^0(\omega)\Bigr]
}.
\label{eq.2.54}
\end{eqnarray}
Here $1+{1 \over 2}[U-g_{\rm r}^2{\hat B}(\omega)]{\hat
  \Pi}^0(\omega)$ and $1+{U \over 2}{\hat \Pi}^0(\omega)$ in the last
line are $3\times 3$-matrices in the $({\hat \rho}_1,{\hat
  \rho}_2,{\hat \rho}_3)$-space introduced in Sec. II. 
Comparing Eq. (\ref{eq.2.54}) with 
Eq. (\ref{eq.2.37}), we see that Bose molecular excitations are the same as the
collective mode resonances of the (generalized) density correlation
functions [unless the denominator in Eq. (\ref{eq.2.54}) diverges]. 
This equivalence is to be expected since the Feshbach coupling in the Hamiltonian
in Eq. (\ref{eq.2.1}) mixes the molecular and Cooper-pair channels.

\newpage

\newpage
%
\begin{figure}
\caption{
\label{fig1} 
Single particle density of states $N(\omega)$ in the weak-coupling BCS region (where the chemical potential $\mu$ is positive). The rapid oscillations in the spectrum are the discrete Bogoliubov quasi-particle energies in a harmonic potential. In this figure, we have introduced a small imaginary part $\Gamma=0.005\omega_0$ to slightly broaden the Bogoliubov quasi-particle peaks. See also Ref. \cite{Ohashi5}.
}
\end{figure}

\newpage

\begin{figure}
\caption{
\label{fig2} 
Spatial variation of the composite order parameter ${\tilde \Delta}(r)$ in the 
BCS region. The parameters are the same as Fig. \ref{fig1}.
The dashed line shows the
``diagonal" potential measured from the chemical potential $\mu$, given by Eq. (\ref{eq.ap2}).
The quasiparticles also feel the Hartree potential $-{U \over 2}n(r)$, but this is
not included.
}
\end{figure}

\centerline{}

\begin{figure}
\caption{
\label{fig3} 
(a) Magnitude of the composite order parameter ${\tilde \Delta}(r=0)$ at the
center of the trap and the single-particle excitation gap $E_g$ in the BCS-BEC
crossover region. (b) Fermi chemical potential $\mu$ in the crossover region. The dotted line in panel (a) shows
the absolute value of the chemical potential $|\mu|$ when $\mu<0$.
}
\end{figure}

\begin{figure}
\caption{
\label{fig4} 
(a) Local density of states $N(\omega,r)$ defined by Eq. (\ref{eq.3.2}) 
in the BCS region ($\nu/\varepsilon_{\rm F}=2$). In this case, 
${\tilde \Delta}(r=0)=10\omega_0\gg\omega_0$ at the center of the trap.
We take ${\bar U}=0.52\varepsilon_{\rm F}$, with the other 
parameters the same as in Figs. \ref{fig1} and \ref{fig2}. To smooth the sharp
delta-functions, we have introduced a small imaginary part $\gamma=0.2\omega_0$ in the Bogoliubov excitation energies. (b) $N(\omega,r)$ based on the LDA. 
In this figure, we take $\gamma=0.1\omega_0$.
The ridge in the LDA single-particle spectrum occurs at $\omega={\tilde \Delta}(r)$.
}
\end{figure}

\begin{figure}
\caption{
\label{fig5} 
(a) Local density of states $N(\omega,r)$ in the BCS region 
($\nu/\varepsilon_{\rm F}=2$) in the case where 
${\tilde \Delta}(r=0)= 1.86\omega_0$ at the center of the trap.
We take ${\bar U}=0.35\varepsilon_{\rm F}$. As usual, the 
finite magnitude of $N(\omega=0,r)$ is due to the small imaginary part
$\gamma=0.2\omega_0$ introduced into the Bogoliubov quasi-particle excitations.
(b) $N(\omega,r)$ in the strong-coupling BEC region, 
where the Fermi chemical potential $\mu$
is negative.
}
\end{figure}

\begin{figure}
\caption{
\label{fig6}
Frequency of the quadrupole mode ($L=2$ and $n=0$) in the BCS-BEC
crossover, for two different values of $\nu$. The threshold energy $2E_{\rm g}$ of two-particle
continuum is evaluated directly from the calculated energy gap in $N(\omega)$, as shown in Fig. 1. 
}
\end{figure}

\begin{figure}
\caption{
\label{fig7}
Effective interaction between molecules mediated by free Fermi atoms in the
BEC limit. The solid and dashed lines represent the Fermi and molecular Bose
Green's functions, respectively.
}
\end{figure}

\begin{figure}
\caption{
\label{fig8}
Spectrum of the strength function 
$S_{33}(\omega,L=2)$ for ${\bar U}=0.35\varepsilon_{\rm F}$. 
The lowest peak corresponds to the quadrupole mode. The arrows show where
the two-particle spectrum ($\omega\ge 2E_g$) starts. 
In this figure, for clarity, we have given the Bogoliubov energies 
a small imaginary part $\gamma=0.005\omega_0$. 
The same peak broadening is also used in Figs.
\ref{fig9}, \ref{fig11}, and \ref{fig12}.
}
\end{figure}

\begin{figure}
\caption{
\label{fig9}
The same as in Fig. \ref{fig8}, for ${\bar U}=0.52\varepsilon_{\rm F}$. Since the two-particle threshold $2E_g\sim 2.5\omega_0$ [see Fig. 6(a)] is much higher than the trap frequency $\omega_0$ in the BCS and crossover region, only the quadrupole mode peak appears in the low energy spectrum.  
}
\end{figure}

\begin{figure}
\caption{
\label{fig10}
Frequency of the monopole mode ($L=0$ and $n=1$) in the BCS-BEC
crossover. $2E_{\rm g}$ is the threshold energy of the two-particle continuum.
}
\end{figure}

\begin{figure}
\caption{
\label{fig11}
Spectrum of the strength function 
$S_{33}(\omega,L=0)$ for ${\bar U}=0.35\varepsilon_{\rm F}$. The 
lowest energy peak corresponds to the monopole mode.
Since the spectral weight of the monopole is very weak in panel (a), 
an inset shows a magnified spectrum around this peak. In panel (a), the
strong peak at $\omega\simeq \omega_0$ comes from two-particle excitations. 
The arrow shows the two-particle threshold energy $2E_{\rm g}$. 
The expected singularity around $\omega=0$ arising from the Goldstone mode
is not shown in this figure.
}
\end{figure}

\begin{figure}
\caption{
\label{fig12}
The same as Fig. \ref{fig11} for ${\bar U}=0.52\varepsilon_{\rm F}$.
}
\end{figure}


\begin{thebibliography}{99}
\bibitem{Jin} C. A. Regal, M. Greiner, and 
D. S. Jin, Phys. Rev. Lett. {\bf 92}, 040403 (2004).
\bibitem{Bartenstein} M. Bartenstein, A. Altmeyer, 
S. Riedl, S. Jochim, C. Chin, 
J. Denschlag, and R. Grimm, Phys. Rev. Lett. {\bf 92}, 120401 (2004).
\bibitem{Zwierlein} M. W. Zwierlein, C. A. Stan, C. H. Schunck, 
S. M. F. Raupach, A. J. Kerman, and W. Ketterle, 
Phys. Rev. Lett. {\bf 92}, 120403 (2004).
\bibitem{Kinast} J. Kinast, S. Hemmer, 
M. Gehm, A. Turlapov, and J. Thomas, 
Phys. Rev. Lett. {\bf 92}, 150402 (2004).
\bibitem{Bourdel} T. Bourdel, L. Khaykovich, K. Cubizolles, 
J. Zhang, F. Chevy, M. Teichmann, L. Tarruell,
S. J. J. M. F. Kokkelmans, and C. Salomon, Phys. Rev. Lett. {\bf 91}, 020402 (2003).
\bibitem{Ohashi1} Y. Ohashi and A. Griffin, Phys. Rev. Lett. {\bf 89},
  130402 (2002).
\bibitem{Ohashi2} Y. Ohashi and A. Griffin, Phys. Rev. A {\bf 67},
  033603 (2003).
\bibitem{Ohashi3} Y. Ohashi and A. Griffin, Phys. Rev. A {\bf 67},
  063612 (2003).
\bibitem{Milstein} J. N. Milstein, S. J. J. M. F. Kokkelmans and
  M. Holland, Phys. Rev. A {\bf 66}, 043604 (2002).
\bibitem{Perali} A. Perali, P. Pieri, and
  G. C. Strinati, Phys. Rev. A {\bf 68}, 031601 (2003).
\bibitem{Eagles} D. M. Eagles, Phys. Rev. {\bf 186}, 456 (1969).
\bibitem{Leggett} A. J. Leggett, in {\it Modern Trend in the Theory of
    Condensed Matter} edited by A. Pekalski and J. Przystawa (Springer
  Verlag, Berlin, 1980), p. 14.
\bibitem{Nozieres} P. Nozi\`eres and S. Schmitt-Rink,
  J. Low. Temp. Phys. {\bf 59}, 195 (1985).
\bibitem{Tokumitsu} A. Tokumitsu, K. Miyake and K. Yamada,
  Phys. Rev. B {\bf 47}, 11988 (1993).
\bibitem{Randeria} M. Randeria, in {\it Bose-Einstein Condensation},
  edited by A. Griffin, D. W. Snoke and S. Stringari (Cambridge
  University Press, N.Y., 1995), p.355.
\bibitem{Melo} C. A. R. S\'a de Melo, M. Randeria and
  J. R. Engelbrecht, Phys. Rev. Lett. {\bf 71}, 3202 (1993).
\bibitem{Engelbrecht} J. R. Engelbrecht, M. Randeria, and C. A. R. Sa
  de Melo, Phys. Rev. B {\bf 55}, 15153 (1997).
\bibitem{Haussmann} R. Haussmann, 
{\it Self-consistent Quantum-field Theory and
Bosonization for Strongly Correlated Electron Systems} 
(Springer-Verlag, Berlin, 1999) Chap. 3.
\bibitem{Timmermans} E. Timmermans, K. Furuya, P. W. Milonni and
  A. K. Kerman, Phys. Lett. A {\bf 285}, 228 (2001).
\bibitem{Holland} M. Holland, S. J. J. M. F. Kokkelmans,
  M. L. Chiofalo and R. Walser, Phys. Rev. Lett. {\bf 87}, 120406
  (2001).
\bibitem{Jin2} T. Loftus, C. A. Regal, C. Tichnor, J. L. Bohn, and
  D. S. Jin, Phys. Rev. Lett. {\bf 88}, 173201 (2002).
\bibitem{Ketterle} K. Dieckmann, C. A. Stan, S. Gupta, Z. Hadzibabic,
  C. H. Schunck, and W. Ketterle, Phys. Rev. Lett. {\bf 89}, 203201
  (2002).
\bibitem{Griffin} A. Griffin, {\it Excitations in a Bose-Condensed Liquid}
(Cambridge University Press, Cambridge, 1993), Chapters 3 and 5.
\bibitem{Chin} C. Chin, M. Bartenstein, A. Altmeyer, S. Riedl, 
S. Jochim, J. Denschlag, and R. Grimm, Science {\bf 305}, 1128 (2004).
\bibitem{Bar2} M. Bartenstein, A. Altmeyer, S. Riedl, S. Jochim, 
C. Chin, J. Denschlag, and R. Grimm, Phys. Rev. Lett. {\bf 92}, 203201 (2004).
\bibitem{OhashiT} Y. Ohashi and S. Takada, J. Phys. Soc. Jpn. {\bf
    66}, 2437 (1997); see also Y. Ohashi and S. Takada, J. Phys. Soc. Jpn. {\bf 67}, 551 (1998).
\bibitem{Ohashi5} Y. Ohashi and A. Griffin, cond-mat/04010220.
\bibitem{Baranov} M. A. Baranov, JETP Lett. {\bf 72}, 385 (2000).
\bibitem{Saint} P. G. de Gennes and D. Saint James, Phys. Lett. {\bf 4}, 151 (1963).
\bibitem{Pethick} See, for example, C. J. Pethick and H. Smith, {\it Bose-Einstein
    Condensation in Dilute Gases} (Cambridge University Press, 2002),
  Chaps. 7 and 14.
\bibitem{BdG} These equations were first studied for the BCS superconductors. See P. G. de Gennes, {\it Superconductivity of Metals and Alloys}
 (Addison-Wesley, N.Y., 1966), Chapter 5; T. Tsuneto, {\it Superconductivity and Superfluidity} (Cambridge University Press, 1998), Chapter 3.
\bibitem{Ohashi4} Y. Ohashi, Phys. Rev. A {\bf 70}, 063613 (2004).
\bibitem{Kohn} W. Kohn, Phys. Rev. {\bf 123}, 1242 (1961); for a recent discussion in the case of a trapped Bose gas, see Sec. 6 of E. Zaremba, T. Nikuni, and A. Griffin, J. Low Temp. Phys. {\bf 116}, 277 (1999).
\bibitem{Bruun1} G. M. Bruun and B. R. Mottelson,
  Phys. Rev. Lett. {\bf 87}, 270403 (2001).
\bibitem{Bruun2} G. M. Bruun, Phys. Rev. Lett. {\bf 89}, 263002 (2002).
\bibitem{Pita} L. Pitaevskii and S. Stringari, Phys. Rev. Lett. {\bf 81}, 4541 (1998).
\bibitem{Stringari} S. Stringari, Europhys. Lett. {\bf 65}, 749 (2004).
\bibitem{Pita2} L. Pitaevskii and S. Stringari, {\it Bose-Einstein Condensation} (Oxford University Press, N.Y., 2003).
\bibitem{Gio7} G. Astrakharchik, J. Boronat, J. Casulleras, S. Giorgini, Phys. Rev. Lett. {\bf 93}, 200404 (2004).
\bibitem{Hei7} H. Heiselberg, Phys. Rev. Lett. {\bf 93}, 040402 (2004), and references given here.
\bibitem{Petrov7} D. S. Petrov, C. Salomon, G. V. Shlyapnikov, Phys. Rev. Lett. {\bf 93}, 090404 (2004).
\bibitem{Ohashi7} Y. Ohashi and A. Griffin, to be published.
\bibitem{Stringari2} S. Stringari, Phys. Rev. Lett. {\bf 77}, 2360 (1996).
\bibitem{note7} In a related context, we note that the standard mean field theory of the BCS-BEC crossover gives rise to a chemical potential which does not agree with the correct Lee-Yang result for a Bose-condensed gas, which is used by Stringari\cite{Stringari}. The better agreement with experimental data which results from using the crossover mean-field expression in the BEC limit [H. Hu, A. Minguzzi, X. Liu and M. Tosi, Phys. Rev. Lett. {\bf 93}, 190403 (2004)] is thus seen to have poor microscopic justification. An improved discussion\cite{Ohashi7} of the BEC limit which includes fluctuations properly must agree with the Lee-Yang equation of state. We thank S. Giorgini for discussion.
\bibitem{Bara5} M. A. Baranov and D. S. Petrov, Phys. Rev. A {\bf 62}, 041601(2000).
\bibitem{Clark} G. M. Bruun and C. W. Clark, Phys. Rev. Lett. {\bf 83}, 5415 (1999).
\bibitem{Ohashi-web} Y. Ohashi, in Workshop on Ultracold Fermi Gases, Levico (Trento), Italy, March 4-6, 2004. This report is available as a pdf-file at: http://bec.science.unitn.it/fermi04.
\bibitem{Ranninger} T. Kostyrko and J. Ranninger, Phys. Rev. B {\bf
    54}, 13105 (1996).
\bibitem{Sch} See for example, J. R. Schrieffer, {\it Theory of Superconductivity} 
(Addison-Wesley, Redwood, California, 1964), Chapter 7.
\bibitem{note2} We subtract the mean-field part from the amplitude 
fluctuations, as ${\hat \rho}_1({\bf r})-\langle{\hat \rho}_1({\bf r})\rangle$.
\bibitem{note3} When we consider the pseudo-spin channel, the 
background attractive interaction $-U$ gives an interaction between
  spin-fluctuations as $H_{\rm FL}^{\rm spin}={U \over 4}{\hat
    s}_z({\bf r}){\hat s}_z({\bf r})$, where ${\hat s}_z({\bf
    r})=\sum_\sigma\sigma\Psi_\sigma^\dagger({\bf r})\Psi_\sigma({\bf
    r})$ is the pseudo-spin density operator. However, since these the spin
  fluctuations do not couple with fluctuations in the density or
  fluctuations in the Cooper pair-channel in GRPA, we have ignored this
  interaction in this paper.
\bibitem{notegg} In Ref. \cite{Ohashi5} and in this paper, we do not include the effect of the fluctuations in renormalizing the value of the order parameter. At $T=0$, this effect is clearly absent but may be more important at finite temperatures, especially in 2D gases. See discussion in J. Keeling, P. Eastham, M. Szymanska, and P. Littlewood, cond-mat/0503184.
\bibitem{CG} See, for example, D. A. Varshalovich, A. N. Moskalev, and V. K. Khersonskii, {\it Quantum Theory of Angular Momentum} (World  Scientific, Singapore, 1988).
\bibitem{Lundh} See for example, E. Lundh, cond-mat/0408569.
\bibitem{notenum} We must take a large cutoff $\omega_c$ to satisfy this condition. This condition becomes crucial when one enters the crossover regime, where the composite order parameter ${\tilde \Delta}(r)$ becomes very large at the center of the trap.  As a result, we need to take a relatively small Feshbach coupling $g_{\rm r}<\varepsilon_{\rm F}$ so that ${\tilde \Delta}(r=0)$ does not exceed $\omega_c$ in the BCS-BEC crossover region.
\bibitem{noteH} Quasi-particles also feel the Hartree diagonal potential $-{U \over 2}n_{\rm F}(r)$, which is taken into account in our calculations.
\bibitem{noteB} The radial component $u^M_{nl}(r)$ has the same form as Eq. (\ref{eq.2.11}), but with the atomic mass $m$ is replaced by the molecular mass $M=2m$. 
\end{thebibliography}
\end{document}